\documentclass[aps,twocolumn,showpacs]{revtex4-1}
\usepackage{bm}
\usepackage{color} 
\usepackage{graphicx}
\usepackage{epsfig}
\usepackage{hyperref}
\usepackage{natbib}
\usepackage{amsmath}
\usepackage{amssymb}

\newcommand{\rmi}{{\rm i}}

\begin{document}

\title{Limitation of electron mobility from hyperfine interaction
in ultra-clean quantum wells and topological insulators}

\author{S.\,A.\,Tarasenko}
\affiliation{Ioffe Institute, 194021 St.~Petersburg, Russia}

\author{Guido Burkard}
\affiliation{Department of Physics, University of Konstanz, D-78464 Konstanz, Germany}

\begin{abstract}
The study of electron transport and scattering processes limiting
electron mobility in high-quality semiconductor structures is central
to solid-state electronics. Here, we uncover an unavoidable source of
electron scattering which is caused by fluctuations of nuclear
spins. We calculate the momentum relaxation time of electrons in
quantum wells governed by the hyperfine interaction between electrons
and nuclei and show that this time drastically depends on the spatial
correlation of nuclear spins. Moreover, the scattering processes
accompanied by a spin flip are a source of the backscattering of Dirac fermions
at conducting surfaces of topological insulators.
\end{abstract}

\pacs{73.50.-h, 73.63.Hs, 71.70.Jp}

\maketitle 

\section{Introduction}

The invention of modulation-doped semiconductor
structures~\cite{Dingle1978} and the subsequent progress in 
semiconductor technology have led to the fabrication of
ultrahigh-mobility two-dimensional 
electron systems~\cite{Umansky2009,Lee2012} and discovery
of novel exciting quantum phenomena such as the fractional quantum
Hall effect~\cite{Prange1990} and microwave-induced resistance oscillations~\cite{Dmitriev2012}. Much effort is focused now on the search for new technological approaches and the optimization of quantum well (QW) design to reduce structure disorder and increase electron mobility. This raises the question of the fundamental limitation of electron mobility that could be achieved in defect-free QW structures with ideal interfaces, see Ref.~\cite{DasSarma2015} for a recent discussion. 

Here, we analyze a source of electron scattering stemming from
hyperfine interaction between electron spins and spins of nuclei
constituting the crystal lattice. This scattering mechanism is
unavoidable in III-V compounds since all stable and long-lived
isotopes of anions (N, P, As, Sb) and cations (B, Al, Ga, In) possess
non-zero nuclear spins. Hyperfine interaction in semiconductors has
been extensively studied in the context of coupled electron and
nuclear spin dynamics in bulk semiconductors~\cite{DP_oo_book},
quantum dots (QDs)~\cite{KKM_spin_book,Urbaszek2013} and
QWs~\cite{Berg1990,Desrat2002,Tifrea2003,Tifrea2011}, and also in the
spin-dependent electron transport along edge channels of a two-dimensional 
electron gas~\cite{Dixon1997,Machida2002,Deviatov2004,LS_spin_book,Maestro2013,Lunde2013} or through QDs 
(spin-blockade effect)~\cite{Ono2002,Johnson2005,Rudner2007,Rudner2010},
but not in bulk charge transport measurements.
Here, we calculate the electron mobility limited by electron-nuclear hyperfine interaction in QWs 
for various spin configurations of the electron and nuclear subsystems
including the case of unpolarized electrons and nuclei and the dynamic
nuclear polarization (DNP). It is shown that the effect of
electron-nuclear interaction on the electron mobility drastically
depends on the spatial correlation of nuclear spins. 
Generally, both spin-conserving and spin-flip processes contribute to
the electron scattering. The quadrupole splitting of the nuclear spin levels in strained QWs or the Zeeman splitting of the electron and nuclear levels in a magnetic field can suppress the spin-flip electron scattering by nuclear spin fluctuations at low temperatures.
A uniform nuclear polarization achieved, e.g., by DNP, suppresses in
turn the spin-conserving scattering processes. Spin-flip scattering 
is an unavoidable source of the backscattering of two-dimensional 
Dirac fermions emerging at conducting surfaces of topological insulators.

\section{Scattering by uncorrelated nuclear spins}

The effective Hamiltonian of hyperfine interaction between the conduction-band electrons and nuclei can be presented in the form~\cite{DP_oo_book}  
\begin{equation}\label{V}
V = \sum_{\alpha,n} A_{\alpha} v_0 \, \bm{S} \cdot \bm{I}_{\alpha,n} \, \delta(\bm{r} - \bm{R}_{\alpha,n}) \:,
\end{equation}
where $\alpha$ is the index of nucleus species, $n$ enumerates nuclei
of certain species, $A_{\alpha}$ are the constants of interaction,
$v_0$ is the volume of the primitive cell, $\bm{S}$ and
$\bm{I}_{\alpha,n}$ are the electron and nucleus spin operators, 
respectively, and $\bm{R}_{\alpha,n}=(\bm{\rho}_{{\alpha,n}},
z_{\alpha,n})$ are the positions of the nuclei~\cite{Footnote1}.

We assume first that nuclei are unpolarized on average and their spin
states are uncorrelated with each other. Then, the scattering of an
electron by different nuclei occurs independently and the total
probability of the scattering from the initial state $(\bm{k},s)$ to
the final state $(\bm{k}',s')$ is given by the golden rule rate
\begin{align}\label{W}
& W_{\bm{k}'s', \bm{k}s} = \frac{2\pi}{\hbar} \sum_{\alpha, n} \sum_{j,j'} |M_{s'j',s j}^{(\alpha,n)}|^2 \, p_{\alpha j} \nonumber \\
& \times \delta\left(\varepsilon_{k} + \varepsilon_s + \varepsilon_{\alpha j} - \varepsilon_{k'} - \varepsilon_{s'} - \varepsilon_{\alpha j'} \right) \,,
\end{align}
where $\bm{k}$ and $\bm{k}'$ are the wave vectors in the QW plane,
$s,s' = \pm 1/2$ are the electron spin projections, $j$ and $j'$ are
the initial and final nuclear spin projections, $M_{s'j',s
  j}^{(\alpha,n)}$ is the matrix element of scattering
at the potential given in Eq.~(\ref{V}), 
$p_{\alpha j}$ is the nuclear spin distribution function,
$\varepsilon_{k} = \hbar^2 k^2 /(2 m^*)$ is the electron kinetic
energy, $m^*$ is the effective mass, $\varepsilon_{s}$ is the energy
related to electron spin, e.g., in an external magnetic field, and $
\varepsilon_{\alpha j}$ are the nuclear energies. 
The hyperfine interaction is weak and short-range as compared to the de Broglie wavelength of electrons. 
It allows us to study the scattering in the first Born approximation and neglect screening.

The squared modulus of the scattering matrix element has the form
\begin{equation}\label{M2}
|M_{s'j',s j}^{(\alpha,n)}|^2 = A_{\alpha}^2 v_0^2 \, |\langle s'j' | \bm{S} \cdot \bm{I}_{\alpha,n} | s j \rangle |^2 \, \psi^4(z_{\alpha,n}) \:,
\end{equation}
where $\psi(z)$ is the function of electron size quantization in the QW. 
The operator $\bm{S} \cdot \bm{I}$ can be rewritten in the form
\begin{equation}
\bm{S} \cdot \bm{I} = S_z I_z + (S_+ I_- + S_- I_+)/2 \,,
\end{equation}
where $S_{\pm} = S_x \pm \rmi S_y$, $I_{\pm} = I_x \pm \rmi I_y$, and 
$S_{\beta}$ and $I_{\beta}$ ($\beta =x,y,z$) are the Cartesian components,
which yields
\begin{align}\label{SI}
& \langle s' j' | \bm{S} \cdot \bm{I} | s j \rangle = s \, j \, \delta_{s's} \delta_{j'j} \nonumber \\
& + \tfrac12 \sqrt{(I-j)(I+j+1)} \, \delta_{s',s-1} \delta_{j',j+1} \nonumber \\
& + \tfrac12 \sqrt{(I-j+1)(I+j)} \, \delta_{s',s+1} \delta_{j',j-1} \,.
\end{align}

The electron relaxation time $\tau_p$ determining the mobility is expressed via the probability of scattering~\eqref{W}.
To derive the expression for $\tau_p$ we follow Boltzmann's approach and write down the collision integral, which plays the role of the ``friction force'' in Boltzmann's equation, 
\begin{equation}\label{St}
{\rm St} f_{\bm k s} = \sum_{\bm k' s'} [f_{\bm k' s'} (1 - f_{\bm k s}) W_{\bm k s, \bm k's'} - f_{\bm k s} (1 - f_{\bm k' s'}) W_{\bm k' s', \bm k s}],
\end{equation}
where $f_{\bm k s}$ is the electron distribution function. In the presence of a weak driving electric field, the distribution function
has the form $f_{\bm k s} = f_{s}^{(0)}(\varepsilon_{k}) + \delta f_{\bm k s}$, where $f_{s}^{(0)}$ is the equilibrium function and $\delta f_{\bm k s}$ is a small anisotropic correction. To first order in $\delta f_{\bm k s}$ and for $W_{\bm k s, \bm k's'}$ independent of the directions of the wave vectors $\bm k$ and $\bm k'$, which is valid for short-range scattering by nuclei, Eq.~\eqref{St} yields
\begin{equation}
{\rm St} f_{\bm k s} = - \frac{\delta f_{\bm k s}}{\tau_p} \,,
\end{equation}
where
\begin{equation}\label{tau1}
\tau_p^{-1}  = \hspace{-0.5mm} \sum_{\bm k' s'} \{ f_{s'}^{(0)}(\varepsilon_{k'}) W_{\bm k s, \bm k's'} + [1 - f_{s'}^{(0)}(\varepsilon_{k'})] W_{\bm k' s', \bm k s} \} . \hspace{-1mm}
\end{equation}

Finally, for the scattering probability given by Eq.~\eqref{W}, homogeneously distributed nuclei and degenerate electron gas, we obtain
\begin{align}\label{tau2}
\tau_p^{-1} & = \frac{m^*}{\hbar^3} \int \psi^4(z) dz \sum_{\alpha} A_{\alpha}^2 v_0^2 N_{\alpha} \sum_{jj's'} |\langle s'j' | \bm{S} \cdot \bm{I}_{\alpha} | sj \rangle|^2  \nonumber \\
& \times [p_{\alpha j} + (p_{\alpha j'} - p_{\alpha j}) f_{s'}^{(0)}(E_F + \varepsilon_{\alpha j} - \varepsilon_{s'}  - \varepsilon_{\alpha j'}) ] \nonumber \\ 
& \times \theta (E_F +  \varepsilon_{\alpha j} - \varepsilon_{s'}  - \varepsilon_{\alpha j'}) \,,
\end{align}
where $N_{\alpha}$ are the densities of nuclei of certain species,
$E_F$ is the Fermi energy, and $\theta(\varepsilon)$ is the
Heaviside step function. We note that, for rectangular QWs with
infinitely high barriers, $\int \psi^4(z) dz = 3/(2d)$ with $d$ being the
QW width. Also note that, in general, $\tau_p$ can depend on the electron spin
$s$ unless the nuclear-spin distribution is symmetric, $p_j = p_{-j}$.
The time reversal symmetry imposing the condition $p_j = p_{-j}$ can be broken
by an external magnetic field or DNP (studied below), 
spontaneously (see discussions on the possibility of nuclear self-polarization in Refs.\cite{DP_oo_book,Abragam_book}),
or, in principle, by driving an electric current through the quantum well. 
In the latter case, the electric current induces a spin polarization of electrons
due to spin-orbit coupling which results, in turn, in a build up of DNP~\cite{Trowbridge2014}. The emerging electron and nuclear spin polarizations are both proportional to the electric current and, therefore, do not affect the linear electron transport. Non-linear effects
are beyond the scope of this paper. 
Below, we analyze Eq.~\eqref{tau2} for some particular cases which can be of interest.  

(i) All spin states of nuclei are degenerate and equally populated, $\varepsilon_{\alpha j}=0$, $p_{\alpha j} = 1/(2I_{\alpha} +1)$, where $I_{\alpha}$ are the nuclear spins; electrons are unpolarized, $\varepsilon_s = 0$. In this case, both spin-conserving and spin-flip scattering processes are allowed. Taking into account that
\[
\sum_{j}  |\langle s \, j | \bm{S} \cdot \bm{I} | s \, j \rangle|^2  = I (I+1)(2I+1)/12 \,,
\]
\[
\sum_{j}  |\langle s \pm 1, j \mp 1 | \bm{S} \cdot \bm{I} | s,j \rangle|^2  = I (I+1)(2I+1)/6 \,,
\]
we obtain
\begin{equation}\label{tau_case1}
\tau_p^{-1} = \frac{m^*}{4\hbar^3} \int  \psi^4(z) dz \sum_{\alpha} A_{\alpha}^2 v_0^2 \, N_{\alpha} \, I_{\alpha} (I_{\alpha} + 1) \,.
\end{equation}
We note that spin-flip scattering processes in the context of electron and nuclear spin relaxation were theoretically considered in Ref.~\onlinecite{Tifrea2003}. 

The degeneracy of nuclear spin levels may be lifted due to the quadrupole interaction of the nuclear spins with the strain-induced gradient of the crystal field in lattice-mismatched structures~\cite{KKM_spin_book}. In a simple axial model relevant for (001)-oriented QWs, the quadrupole interaction determining the splitting and order of the spin levels is proportional to $I_z^2$, i.e., levels with different $j^2$ have different energies. The sign of the quadrupole splitting is opposite for axial tension and compression.
Typical values of the quadrupole splitting in III-V heterostructures are of the order of $1$ -- $10$\,neV which corresponds to $10^{-5}$ -- $10^{-4}$\,K at the temperature scale~\cite{KKM_spin_book,Kuznetsova2014}. Therefore, non-equal thermal population of the nuclear spin levels at sub-mK temperatures can occur and affect the electron-nuclear interaction.

(ii) Nuclear spin levels are split by strain in such a way that the
ground levels are characterized by the highest spin projections $j =
\pm I_{\alpha}$, $I_{\alpha} > 1/2$; electrons are unpolarized, $\varepsilon_s = 0$. Temperature is lower than the
nuclear quadrupole splitting and, therefore, only ground levels
contribute to scattering, $p_{\alpha j} = 1/2$ if $j = \pm I_{\alpha}$
and $p_{\alpha j} = 0$ otherwise.  Only spin-conserving scattering 
processes can occur and the corresponding electron relaxation time is given by 
\begin{equation}\label{tau_case3}
\tau_p^{-1} =  \frac{m^*}{4\hbar^3} \int \psi^4(z) dz \sum_{\alpha} A_{\alpha}^2 v_0^2 \, N_{\alpha} I_{\alpha}^2 \,.
\end{equation}

(iii) Nuclear spin levels are split by strain in such a way that the ground levels are characterized by the lowest spin projections $j = \pm 1/2$, $I_{\alpha}$ is half-integer; electrons are unpolarized, $\varepsilon_s = 0$. Temperature is lower than the quadrupole splitting energy and only ground levels contribute to scattering, $p_{\alpha, \pm 1/2} = 1/2$. In this case, both spin-flip and spin-conserving processes contribute to scattering and the electron relaxation time has the form 
\begin{equation}\label{tau_case4}
\tau_p^{-1} =  \frac{m^*}{8\hbar^3} \int \psi^4(z) dz \sum_{\alpha} A_{\alpha}^2 v_0^2 \, N_{\alpha} [ 1/2 + (I_{\alpha}+1/2)^2 ] \,.
\end{equation}

(iv) All spin states of nuclei are degenerate and equally populated, $\varepsilon_{\alpha j}=0$, $p_{\alpha j} = 1/(2I_{\alpha} +1)$; electrons are completely spin polarized by an external magnetic field (thermal nuclear polarization is small and neglected). In this particular case, only spin-conserving processes can occur and Eq.~\eqref{tau2} yields
\begin{equation}\label{tau_case2}
\tau_p^{-1} = \frac{m^*}{12 \hbar^3} \int \psi^4(z) dz \sum_{\alpha} A_{\alpha}^2 v_0^2 \, N_{\alpha} \, I_{\alpha} (I_{\alpha} + 1) \,.
\end{equation}

In III-V semiconductor structures, nuclear spins can be efficiently polarized by DNP~\cite{
DP_oo_book,KKM_spin_book}. In the case of uniform nuclear polarization,
the average nuclear field should be excluded from the Hamiltonian of
electron-nuclear interaction causing the electron scattering because
it does not introduce any disorder which breaks the translational symmetry of the crystal lattice. Accordingly, the operators $\bm{I}_{\alpha}$ in Eqs.~\eqref{M2} and~\eqref{tau2} should be replaced by 
$\bm I_{\alpha} - \tilde{\bm I}_{\alpha}$, where the vectors $\tilde{\bm I}_{\alpha}$ are given by
\begin{equation}
\tilde{\bm I}_{\alpha} = \frac{\sum_{\beta} A_{\beta} N_{\beta} \bar{\bm I}_{\beta}}{A_{\alpha} \sum_{\beta} N_{\beta}} \,,
\end{equation}
the index $\beta$ runs over the nuclei of anions or cations if $\alpha$ stands for a nucleus of an anion or a cation, respectively,
and $\bar{\bm I}_{\alpha}$ is the average nuclear spin of a certain isotope. In the simple case of an isotopically pure crystal, where all anions and cations are of certain isotopes, $\tilde{\bm I}_{\alpha} = \bar{\bm I}_{\alpha}$. We assume that all the vectors $\bar{\bm I}_{\alpha}$ point along the same axis $z$ and choose $z$ as the spin quantization axis. Then, one obtains $\bar{I}_{\alpha} = \bar{j}_{\alpha} = \sum_{j} j p_{\alpha j}$ and the scattering marix elements
\begin{align}
& \langle s' j' | \bm{S} \cdot ( \bm{I}_{\alpha} - \tilde{\bm I}_{\alpha} ) | s j \rangle = s \, (j - \tilde{I}_{\alpha} ) \,\delta_{s's} \delta_{j'j} \nonumber \\
& + \tfrac12 \sqrt{(I_{\alpha} - j)(I_{\alpha} + j + 1)} \, \delta_{s',s-1} \delta_{j',j+1} \nonumber \\
& + \tfrac12 \sqrt{(I_{\alpha} - j + 1)(I_{\alpha} + j)} \, \delta_{s',s+1} \delta_{j',j-1} \,.
\end{align}
Now we discuss the electron relaxation time in the presence of DNP.

(v) The nuclei are spin polarized; electrons of both spin states are present at the Fermi level.
The splitting of nuclear levels is lower than temperature so that both spin-conserving and spin-flip processes can contribute to scattering. In such conditions, the momentum relaxation time becomes spin dependent and is given by
\begin{equation}\label{tau_case5}
\tau_{p,s}^{-1} = \frac{m^*}{4\hbar^3}  \hspace{-1mm} \int  \hspace{-1mm} \psi^4(z) dz  \hspace{-1mm} \sum_{\alpha}
A_{\alpha}^2 v_0^2 N_{\alpha} [ I_{\alpha}^2 + I_{\alpha}  + \tilde{I}_{\alpha}^{\,2} - 2 \bar{j}_{\alpha} (\tilde{I}_{\alpha} + s) ] .
\end{equation}

(vi)  The nuclei are spin polarized; electrons are completely spin polarized by the nuclear (Overhauser) effective magnetic field or an external magnetic field. Then, only spin-conserving scattering occur and we 
obtain the relaxation time 
\begin{equation}\label{tau_case6}
\tau_p^{-1} = \frac{m^*}{4\hbar^3} \int  \psi^4(z) dz \sum_{\alpha}
A_{\alpha}^2 v_0^2 \, N_{\alpha} [ \, \overline{j_{\alpha}^2} - 2 \bar{j}_{\alpha} \tilde{I}_{\alpha} + \tilde{I}_{\alpha}^{\,2}]\, ,
\end{equation}
where $\overline{j_{\alpha}^2} = \sum_j j^2 \, p_{\alpha j}$. For isotopically purified crystals, the expression in the square brackets is reduced to ${\overline{j_{\alpha}^2} - \bar{j}_{\alpha}^{\,2}}$. This quantity is zero for fully polarized nuclei when both $\overline{j_{\alpha}^2}$ and $\bar{j}_{\alpha}^{\,2}$ are equal to $I_{\alpha}^{2}$. The absence of spin-conserving electron scattering by the fully polarized nuclei, when the crystal translational symmetry is restored, is in accordance with the free motion of Bloch electrons in a periodic potential. In isotopically mixed QW structures, the scattering may occur even in the case of fully polarized nuclei due to the disorder caused by a difference in the nuclear spins $I_{\alpha}$ and/or the interaction constants $A_{\alpha}$ of anions (or cations). The corresponding relaxation time is given by  
\begin{equation}\label{tau_case6b}
\tau_p^{-1} = \frac{m^*}{4\hbar^3} \int  \psi^4(z) dz \sum_{\alpha}
A_{\alpha}^2 v_0^2 \, N_{\alpha} (I_{\alpha} - \tilde{I}_{\alpha})^2 \,.
\end{equation}

Now we estimate the electron relaxation time governed by the scattering from unpolarized nuclei following Eq.~\eqref{tau_case1}. The estimation for a 10-nm-wide QW grown from GaAs, where $m^* \approx 0.067 m_0$ with $m_0$ being the free electron mass, $I_{{\rm Ga}}=3/2$, $I_{{\rm As}} = 3/2$, and $v_0^2 \sum_{\alpha} A_{\alpha}^2 N_{\alpha} \approx 0.2 \times 10^{-24}$ meV$^2$\,cm$^3$ (Ref.~\onlinecite{Merkulov2002}), gives $\tau_p \sim 10^{-5}\,$s. 
This corresponds to the mobility $\mu \sim 4 \times 10^{11}$ cm$^2$/(Vs) that is still a few orders of magnitude higher than the mobility achieved to date~\cite{Umansky2009}. The relaxation time governed by the hyperfine interaction can be much shorter in structures made of atoms with large nuclear spins (e.g., $I_{{\rm In}}=9/2$) or heavy atoms, where the interaction constants are larger. 

The probabilities of spin-conserving and spin-flip processes are comparable.
Therefore, electron scattering by unpolarized nuclei makes a contribution to electron spin relaxation with the time $\tau_s$ comparable to the momentum relaxation time $\tau_p$ calculated above. This spin relaxation mechanism can be important if other mechanisms are
suppressed, e.g., in (110)-oriented QWs where the D'yakonov-Perel' spin relaxation mechanisms is suppressed for the out-of-plane spin component~\cite{Dyakonov86,Tarasenko2009,Poshakinskiy2013} and the spin lifetime up to $0.5$$\,\mu$s has been recently determined~\cite{Voelkl2014}.

\section{Scattering of Dirac fermions}

Spin-flip processes contribute to the backscattering of two-dimensional Dirac fermions emerging at surfaces of 3D topological insulators (TIs). In such materials, the strong spin-orbit interaction giving rise to topologically protected surface states locks the carrier spin and momentum~\cite{Hasan2010}. Therefore, elastic scattering between the states with the opposite momenta is forbidden in the presence of time reversal symmetry. Interaction with nuclear spins breaks the time reversal symmetry in the subsystem of the Dirac fermions and enables the backscattering. Electron-nuclear interaction leading to backscattering between one-dimensional helical edge states of 2D TIs was studied in Refs.~\onlinecite{Maestro2013,Lunde2013}. Here, we calculate the probability of scattering for two-dimensional Dirac fermions on surfaces of 3D TIs and provide estimates for HgTe and Bi$_2$Se$_3$.

HgTe is a gapless semiconductor with the inverted band structure which becomes a 3D TI if strained and the strain opens a gap in the otherwise four-fold degenerate $\Gamma_8$ states~\cite{Bruene2010,Dantscher2015}. Within the 6-band $\bm k$$\cdot$$\bm p$ theory, relevant for narrow-band materials, the topological surface states are described by the wave functions
\begin{align}\label{Psi_topins}
\Psi_{\bm k}(\bm \rho,z) & = \left[ \sum_{m = \pm 1/2} \psi_{\Gamma_6, m}(z) |\Gamma_6, m \rangle \right. \\
& \left. + \sum_{m = \pm 1/2, \pm 3/2} \psi_{\Gamma_8, m}(z) |\Gamma_8 ,
  m \rangle \right] \exp( i \bm k \cdot \bm \rho) \,, \nonumber
\end{align}
where $\psi_{\Gamma_6, m}(z)$ and $\psi_{\Gamma_8, m}(z)$ are the envelope
functions in the direction normal to the surface, $|\Gamma_6, m \rangle$ and $|\Gamma_8, m \rangle$ are the
the basis Bloch amplitudes of the $s$-type $\Gamma_6$ and $p$-type
$\Gamma_8$ states, respectively, and $\bm k$ is the wave vector in the
surface plane. For strained HgTe films, the wave
function~\eqref{Psi_topins} contains considerable contribution
($\sim$20 \%) of the $\Gamma_6$ states~\cite{Dantscher2015}. Since the
hyperfine interaction for $s$-type Bloch amplitudes given by the Fermi
contact term is much stronger than
that for $p$-type Bloch amplitudes~\cite{KKM_spin_book,Gryncharova1977}, a good estimation is that the
electron-nuclear interaction for the Dirac fermions at HgTe surface is
of contact type and determined by the contribution of the $\Gamma_6$ states.

The electron spin of the surface states in the axial approximation lies in the surface plane and points perpendicular to the wave vector. Accordingly, the spinor composed of the functions $\psi_{\Gamma_6, \pm 1/2}(z)$ can be presented in the form 
\begin{equation}
\left[
\begin{array}{c}
\psi_{\Gamma_6, +1/2}(z) \\
\psi_{\Gamma_6, -1/2}(z)
\end{array}
\right] 
= \frac{1}{\sqrt{2}} 
\left[
\begin{array}{c}
1 \\
i \exp(i\varphi_{\bm k})
\end{array}
\right] 
\psi_{\Gamma_6}(z) \,,
\end{equation}
where $\varphi_{\bm k}$ is the polar angle of the vector $\bm k$. Following Eqs.~\eqref{V} and~\eqref{SI} one can readily calculate the
matrix element of scattering. For the scattering from the initial state $\bm k$ to the final state $\bm k'$ the squared modulus of the scattering matrix element has the form
\begin{align}
|M_{\bm k' j', \bm k j}^{(\alpha,n)}|^2 =  \frac{1}{16} A_{\alpha}^2 v_0^2 \, |\psi_{\Gamma_6}(z_{\alpha,n})|^4 \left[ 
4 \sin^2 (\theta/2) \, j^2 \delta_{j'j} \;\;\;\;\;\;\;\; \right. \nonumber \\
 + (I_{\alpha}-j+1)(I_{\alpha}+j)  \delta_{j',j-1} +   (I_{\alpha}-j)(I_{\alpha}+j+1)  \delta_{j',j+1} ], \nonumber 
\end{align}
where $\theta = \varphi_{\bm k'} - \varphi_{\bm k}$ is the angle of scattering. 

The differential probability of elastic scattering of a test particle by the angle $\theta$ in 2D systems is given by
\begin{equation}
d w(\theta) = \frac{k}{2 \pi \hbar^2 v} \sum_{\alpha, n} \sum_{j,j'} |M_{\bm k' j', \bm k j}^{(\alpha,n)}|^2 p_{\alpha j} \, d\theta \,,
\end{equation}
where $v = (1/\hbar) d \varepsilon_k / d k$ is the velocity which is independent of the energy for linearly dispersive Dirac fermions.

We assume that all spin states of the nuclei are degenerate and
equally populated, $p_{\alpha j} = 1/(2I_{\alpha} +1)$. 
Then, summing up over the nuclei, we obtain
\begin{align}\label{w_DF}
d w(\theta) &= \frac{k}{24 \pi \hbar^2 v}  \left[ 1 +  \sin^2 (\theta/2) \right] d \theta \nonumber \\
&\times \int |\psi_{\Gamma_6}(z)|^4 dz \sum_{\alpha} A_{\alpha}^2 v_0^2 \, N_{\alpha} \, I_{\alpha} (I_{\alpha} + 1) \,.
\end{align}
Equation~\eqref{w_DF} describes the scattering of two-dimensional Dirac fermions by nuclear spin fluctuations. It shows that the backscattering ($\theta = \pi$) determined by spin-flip processes is twice as efficient as the forward scattering ($\theta = 0$) determined by spin-conserving processes.

In natural HgTe, about 30~\% of Hg nuclei and about 8~\% of Te nuclei possess non-zero spins~\cite{Lunde2013}. To the best of our knowledge, the hyperfine interaction constants have not been measured yet. Considering the fraction of nuclei with non-zero spins, the typical parameters of surface states in HgTe films~\cite{Dantscher2015}: $k = 2 \times 10^6$~cm$^{-1}$, $v = 0.5 \times 10^8$~cm/s, the characteristic length of the surface-state localization $d = 10$~nm, and the $\Gamma_6$ band partition $0.2$, we estimate that the probability of scattering by nuclear spin fluctuations is two-three orders of magnitude lower than that in GaAs quantum wells.

Other prominent examples of 3D TIs are binary and ternary compounds of Bi with Se and Te~\cite{Hasan2010}. Natural Bi consists of the only isotope $^{209}$Bi with the nuclear spin $9/2$. Recent measurements of nuclear magnetic resonance in $n$-type Bi$_2$Se$_3$ crystals have revealed very strong contact interaction between electrons and $^{209}$Bi nuclei (in spite of the fact that the Bloch amplitude is mostly of $p$-type)~\cite{Mukhopadhyay2015}. The contact hyperfine interaction constant is found to be comparable and even exceed those for the conduction-band electrons in GaAs. These results suggest that the scattering of Dirac fermions at the Bi$_2$Se$_3$ surface by nuclear spin fluctuations can be as efficient as that for electrons in GaAs-based structures.

\section{Scattering by fluctuations of macroscopic nuclear polarization}

The electron-nuclear interaction is drastically enhanced if the nuclear spins are spatially correlated and polarized at a macroscopic scale. Such a nuclear spin polarization inhomogeneous in the QW plane can be created, e.g., via DNP by optical grating technique~\cite{Cameron1996,Weber2007,Wang2013}. In the case of macroscopic nuclear polarization, the interaction can be described as the Zeeman term  
\begin{equation}\label{V_macr}
V(\bm{\rho}) = g \mu_B \, \bm{S} \cdot \bm{B}_n(\bm{\rho}) 
\end{equation}
with the effective nuclear (Overhauser) magnetic field 
\begin{equation}\label{B_n}
\bm{B}_n(\bm{\rho}) = \sum_{\alpha} (A_{\alpha} v_0 N_{\alpha}/ g \mu_B) \int \overline{ \bm{I}_{\alpha}(\bm{\rho},z) } \, \psi^2(z) dz 
\end{equation}
which varies in the QW plane at a scale much larger than the crystal lattice constant. Here, $\mu_B$ is the Bohr magneton, $g$ is the effective electron $g$-factor, and the overline denotes quantum mechanical averaging over the ensemble of nuclear wave functions.

The effective magnetic field Eq.~\eqref{V_macr} produces a spin-dependent electron potential which causes the scattering. 
For the effective field $\bm{B}_n(\bm{\rho})$ oriented along a certain axis, e.g., the growth direction, the momentum relaxation time assumes the form 
\begin{equation}\label{tau_macr}
\tau_p^{-1} = \frac{m^* (g \mu_B)^2}{8 \pi \hbar^3} \hspace{-1mm} \int_{-\pi}^{\pi} \hspace{-2mm} \langle \bm B_n(\bm{\rho}) \cdot \bm B_n(\bm{\rho}') \rangle_{2k_F \sin \theta /2} 
(1-\cos\theta) d\theta \:,
\end{equation}
where $\langle \bm B_n(\bm{\rho}) \cdot \bm B_n(\bm{\rho}') \rangle_{\bm{q}}$ is the Fourier image of the spatial correlation function 
$\langle \bm B_n(\bm{\rho}) \cdot \bm B_n(\bm{\rho}') \rangle$, $k_F$ is the Fermi wave vector, and $\theta$ is the angle of scattering.

To estimate the momentum relaxation time we assume that the nuclear polarization is randomly distributed in the QW plane with the characteristic correlation length~$l$ and zero mean value. Then, for GaAs-based QWs with the nuclear polarization $\overline{\bm{I}_{\alpha}} \sim 1$, $v_0 \sum_{\alpha} A_{\alpha} N_{\alpha} \approx 0.1$ meV (Ref.~\onlinecite{Merkulov2002}), the Fermi wave vector 
$k_F = 10^6$\,cm$^{-1}$, and the correlation length $l \sim 1/k_F$, one obtains $\tau_p \sim 10^{-9}$ s.
Such $\tau_p$ is comparable to the momentum relaxation time in
high-mobility structures. It indicates that strong spatially
inhomogeneous spin polarization of nuclei can considerably affect the
electron transport in quantum wells and can be probed by electrical
measurements. Spatially oscillating nuclear polarization created by optical grating
technique may cause the Bragg diffraction of electrons and has even stronger impact on the electron transport.

To summarize, we have calculated the limitation of electron mobility in III-V quantum wells from the unavoidable source of disorder originating from the fluctuations of nuclear spins. We have analyzed various spin configurations of the electron and nuclear subsystems
and shown that the electron mobility determined by the electron-nuclear hyperfine interaction drastically depends on the spatial correlation of nuclear spins. While the electron mobility limited by the hyperfine interaction with uncorrelated nuclear spins is still few orders of magnitude higher than that achieved in high-mobility quantum wells, nuclear spins that are spatially correlated and polarized at a macroscopic scale can considerably affect the electron transport in modern high-mobility quantum wells.

\section*{Acknowledgments}

The work of SAT was supported by the Russian Science Foundation (project no. 14-12-01067),
GB received support from DFG within the SFB767.


\begin{thebibliography}{99}
\bibitem{Footnote1} We consider a simple model and neglect terms which may arise due to structure anisotropy.
\end{thebibliography}
\end{document}